\documentclass[prl,aps,twocolumn,showpacs,floatfix,superscriptaddress]{revtex4}
\usepackage{epsfig}
\usepackage{amsmath}
\usepackage{amssymb}
\usepackage{amsfonts}
\usepackage{mathptmx}
\usepackage{eucal}

\newcommand{\lam}{\lambda_\perp}
\newcommand{\Ic}{I_c^{\textit{GL}}}
\newcommand{\IA}{I_c^{\textit{AL}}}
\begin{document}

\title{Advances in the criteria for dividing thin superconducting films into narrow
and wide films}

\author{V. M. Dmitriev}
\affiliation{B.Verkin Institute for Low Temperature Physics and Engineering,
Kharkov 61103, Ukraine}
\affiliation{International Laboratory of High Magnetic Fields and Low
Temperatures, Wroclaw 53-421, Poland}
\author{I. V. Zolochevskii}
\email{zolochevskii@ilt.kharkov.ua}
\affiliation{B.Verkin Institute for Low Temperature Physics and Engineering,
Kharkov 61103, Ukraine}
\author{E. V. Bezuglyi}
\affiliation{B.Verkin Institute for Low Temperature Physics and Engineering,
Kharkov 61103, Ukraine}

\begin{abstract}
The results of experimental investigations of the critical currents and
certain nonequilibrium phenomena in thin Sn films of different width $w$ are
analyzed. Usually, thin superconducting films are divided into two groups:
narrow channels $w<\lam$ and wide films $w>\lam$. A wide transitional region
where the condition $w>\lam$ holds with a large margin and at the same time
cannot be explained from the standpoint of the theory of the appearance of a
vortex state has been found. This shows that the generally accepted criterion
$w\sim \lam$ for dividing films into wide and narrow does not work. The
transition into a wide-film regime, described by the existing theory of the
vortex state, is fully completed only for $w/\lam(T) > 10-20$.
\end{abstract}


\maketitle

It is now generally accepted that thin superconducting films can be divided
into narrow and wide films. Films whose width $w$ and thickness $d$ are
smaller than the penetration depth $\lam(T)$ of a magnetic field and the
coherence length $\xi(T)$, which ensures a uniform distribution of the
current over the film width and absence of vortex resistivity, are said to be
narrow channels. Phase-slip centers (PSCs) appear when the current, exceeding
the critical current $\Ic (T)$ for Ginzburg-Landau (GL) depairing, flows in a
narrow channel. These centers consist of the cores of size $\sim\xi(T)$ and
the diffusion "tails" on both sides of the core, having the length of the
penetration depth of a longitudinal electric field into the superconductor.
The order parameter $\Delta$ and the superconducting current $I_s$ in the
core of a PSC oscillate, and at certain moments $\Delta$ and $I_s$ vanish and
the phase changes abruptly by $2\pi$. The frequency of the oscillations is
determined by the Josephson ratio $f_J =2 eV/h$.

Films whose width $w>\lam(T),\xi(T)$ are said to be wide \cite{1}. In such
films, due to the Meissner screening of the cur\-rent-induced magnetic field,
the current distribution over the width is nonuniform and exhibits sharp
peaks at the edges. According to the Aslamazov-Lempitskiy (AL) theory
\cite{2}, the resistive transition of a wide film is due to the instability
of the current state when the edge current density reaches a value close to
the critical current density $j_c^ {GL}$ in the GL theory. This instability
results in the creation of vortices, whose motion generates voltage at the
ends of the film. The corresponding critical current $\IA$ is small compared
to $\Ic$, since the current density in the bulk of the film is far from the
critical value. On account of the motion and annihilation of the vortices, a
second sharp peak in the current density forms at the center of the film. For
certain value $I_m$ of the current, the magnitude of this peak reaches a
critical value $j_c^ {GL}$, which results in instability of the stationary
flow of the vortices. In the AL theory, it was supposed that for $I>I_m$ a
film exhibits transition into the normal state.

\begin{figure}[tb]
\centerline{\epsfxsize=8cm\epsffile{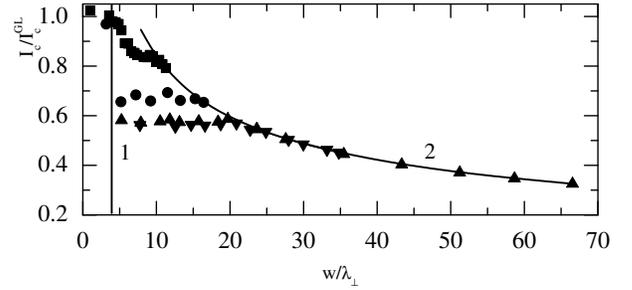}} \vspace{-3mm}
\caption{Reduced critical current density $I_c / \Ic$ versus $w/\lam$ for the
samples SnW5 ($\blacktriangle$), SnW6 ({\Large $\bullet$}), SnW8
($\blacktriangledown$), SnW9 ($\blacklozenge$), SnW10 ($\blacksquare$); curve
2 -- AL theoretical curve. The parameters of the films can be found in Ref.
4.}
\label{fig1}\vspace{-3mm}
\end{figure}

The experimental investigations of resistive transitions in wide films,
although qualitatively confirming the above-des\-cribed picture of a
nonuniform distribution of the current in subcritical and vortex states
\cite{3,4}, have at the same time resulted in considerable refinement of the
main assumptions of AL. It turned out \cite{4} that the vortex resistivity
arises only in a film of sufficient width, $w>4\lam$, while for $I>I_m$ the
film passes not into a normal state but rather to a vortex-free state with
phase-slip lines (PSLs) -- the two-dimensional analogues of PSCs. Even more
unexpected, and prompting us to raise the question of the true criterion for
a ``wide film'', was the behavior of the temperature dependence of the
critical current, shown in Fig.~1 for several samples as a dependence of the
reduced current $I/ \Ic $ on the parameter $w/\lam$, which is proportional to
$1-T/T_c$. According to the AL theory, for $w>\lam$ this dependence should be
described by a universal function of $w/\lam$, $\IA/ \Ic
=1.5(\pi\lam/w)^{1/2}$ (curve 2). In the experiment, as seen from Fig.~1,
this universality does indeed exist but only for very large values of $w/\lam
\sim 10-20$, which therefore should serve as the real criterion for a
transition into the wide-film regime.

For smaller values of $w/\lam$, the behavior of the critical current is very
peculiar. According to Ref.~4, samples with $w/\lam < 4$ are narrow channels
(in Fig.~1 the region to the left of the vertical straight line 1). In the
interval between the vertical straight line 1 and the curve 2, there is a
transitional region where the film is ``quasi-wide'' (a vortex section is
present in the current-voltage characteristic (IVC) but the AL theory ``does
not work'' yet). For the SnW10 film ($w=7 \;\mu$m) the experimental current
$I_c$ for $w/\lam <4$ equals the computed value of $\Ic$. As temperature
decreases, i.e. $w/\lam$ increases, vortex resistivity arises and $I_c$ drops
rapidly to a smaller value, which nonetheless reveals the temperature
dependence close to $I_c(T)$ that continues right up to a transition into
regime described by the AL theory. This decrease of $I_c$ is even more rapid
for the SnW6 film ($w=17\;\mu$m), as a result of which the region of the
transitional temperature dependence $I_c \propto (1-T/T_c)^{3/2}$ expands
considerably. Finally, for the widest films SnW8 ($w =25\;\mu$m) and SnW5
($w=42\;\mu$m), where a narrow-channel regime is not observed because of the
extreme proximity of the corresponding temperature region to $T_c$, the
dependencies of $I_c / \Ic$ on $w/\lam$ are completely identical, and the
width of the transitional region reaches saturation.

It should be noted that the observed wide transitional region, where the
condition $w>\lam$ holds with a large margin and at the same time the
dependence $I_c(T)$ is identical to the behavior of $\Ic(T)$, cannot be
explained from the standpoint of the AL theory of the mechanism of the
appearance of the vortex state. A characteristic feature of this region is a
sharp transition of $I_c\propto (1-T/T_c)^{3/2}$ to $\IA(T) \propto 1-T/T_c$.
Our numerical solution of the London integral equation \cite{2} for the
current distribution over the film width for arbitrary values of $w/\lam$
also shows the absence of a wide transitional region between $\Ic(T)$ and
$\IA(T)$ provided the critical current density $j_c^{ GL}$ is reached at the
edge of the film at the point of the transition. This could signify that the
condition for the appearance of vortices for $w/\lam <10-20$ weakens because
of the existence of a competing mechanism for the penetration of vortices
into the film when the edge current density reaches some value $j_c(T)$
smaller than $j_c^ {GL}(T)$. To explain the observed constancy of the value
of $Ic / \Ic$ in the transitional region, it must be assumed that as the
temperature decreases the quantity $j_c(T)$ increases more rapidly than $j_c^
{GL}(T)$, and is proportional to $(1-T/T_c)^2$. At the point where they
become equal, a transition occurs into the AL regime, which explains the
sharpness of the crossover to the linear temperature dependence of $\IA(T)$
seen in Fig. 1. In principle, such a mechanism can be associated with edge
microdefects, which have virtually no effect on the GL depairing current in
the narrow-film regime, but they are a source of vortices in a wide film. At
the same time the excellent quantitative agreement between the critical
currents $I_c(T)$, $I_m(T)$ and the predictions of the AL theory for large
values of $w/\lam$ casts doubt on this supposition.

\begin{figure}[tb]
\centerline{\epsfxsize=8cm\epsffile{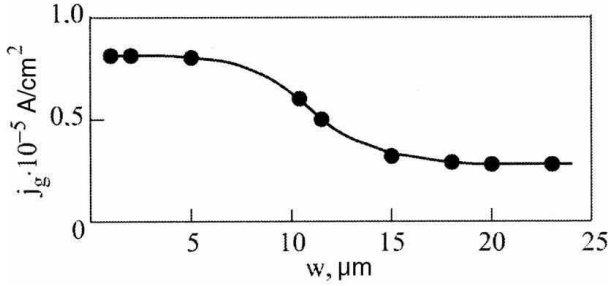}} \vspace{-3mm}
\caption{Generation current density versus Sn film width at $T/T_c=0.99$.}
\label{fig2}\vspace{-3mm}
\end{figure}

Other nonequilibrium phenomena, for example, the generation of
electromagnetic oscillations with frequency lower than the Josephson
frequency, also point to the inapplicability of the widely accepted criterion
$w \approx \lam(T)$ for a ``wide film''. This phenomenon, called
non-Josephson generation \cite{5,6}, arises when a certain current density
$j_g$ is reached, and this generation current density for narrow tin films
($w\approx \xi \approx 1 \;\mu$m) is 2.7 times higher than for wide samples
with $w >20 \;\mu$m at the same reduced temperature $T/T_c=0.99$ \cite{6}.
Thus, for film width in the range $1-20 \;\mu$m, a smooth transition can
occur from high to low generation current densities. To check this
assumption, a film obtained in one deposition operation was used to prepare a
series of samples with different width. The results of measurements of the
generation current density $j_g$ versus the width $w$ of tin films at
temperature $T=0.99T_c$ \cite{6} are presented in Fig. 2. It is evident that
the center of the transition from a narrow to a wide channel corresponds to
$w\approx 12 \;\mu$m. Thus, from the standpoint of non-Josephson generation,
tin films with width up to $w\approx 5 \;\mu$m can be treated as narrow
channels, while they become ``wide'' only for $w>20 \;\mu$m.

A similar conclusion concerning the inapplicability of the widely accepted
criterion $w\approx \lam(T)$ \cite{1} for determining the boundary between
narrow and wide films can also be drawn on the basis of an analysis of
changes in the dependence of the critical current $I_c$ on the power $P$ of
an external microwave field for films with different width \cite{7}. Figure
3a shows the function $I_c(P)$ in relative units for tin films of different
width under identical experimental conditions. The descending sections of the
curves $I_c(P) / I_c(0)$ in Fig. 3a can be fit by the function $I_c \propto
(1-P/P_c)^\alpha$. For the sample SnW10 ($w=7 \;\mu$m), which at $T/T_c=0.99$
is a narrow channel \cite{4}, the dependence $I_c(P) / I_c(0)$ is convex and,
correspondingly, $\alpha = 0.53<1$. For the sample SnW6 ($w=17 \;\mu$m) this
dependence becomes concave and $\alpha =1.4$ is larger than 1, while for the
obviously wide film SnW5 ($w=42 \;\mu$m) $\alpha =2.9$. The dependence of the
exponent of the fitting functions of the critical current versus the sample
width, presented in Fig. 3b, shows that the value $\alpha =1$ can be expected
for a film of width $w\approx 12 \;\mu$m, and the descending section of the
function $I_c(P)$ is a straight line. In other words, as the film width
increases, the sign of the curvature of the descending sections of the curves
changes, and a film with $w\approx 12 \;\mu$m is a boundary between
quasi-narrow and quasi-wide samples, as in the case displayed in Fig. 2.

\begin{figure}[th]
\centerline{\epsfxsize=8cm\epsffile{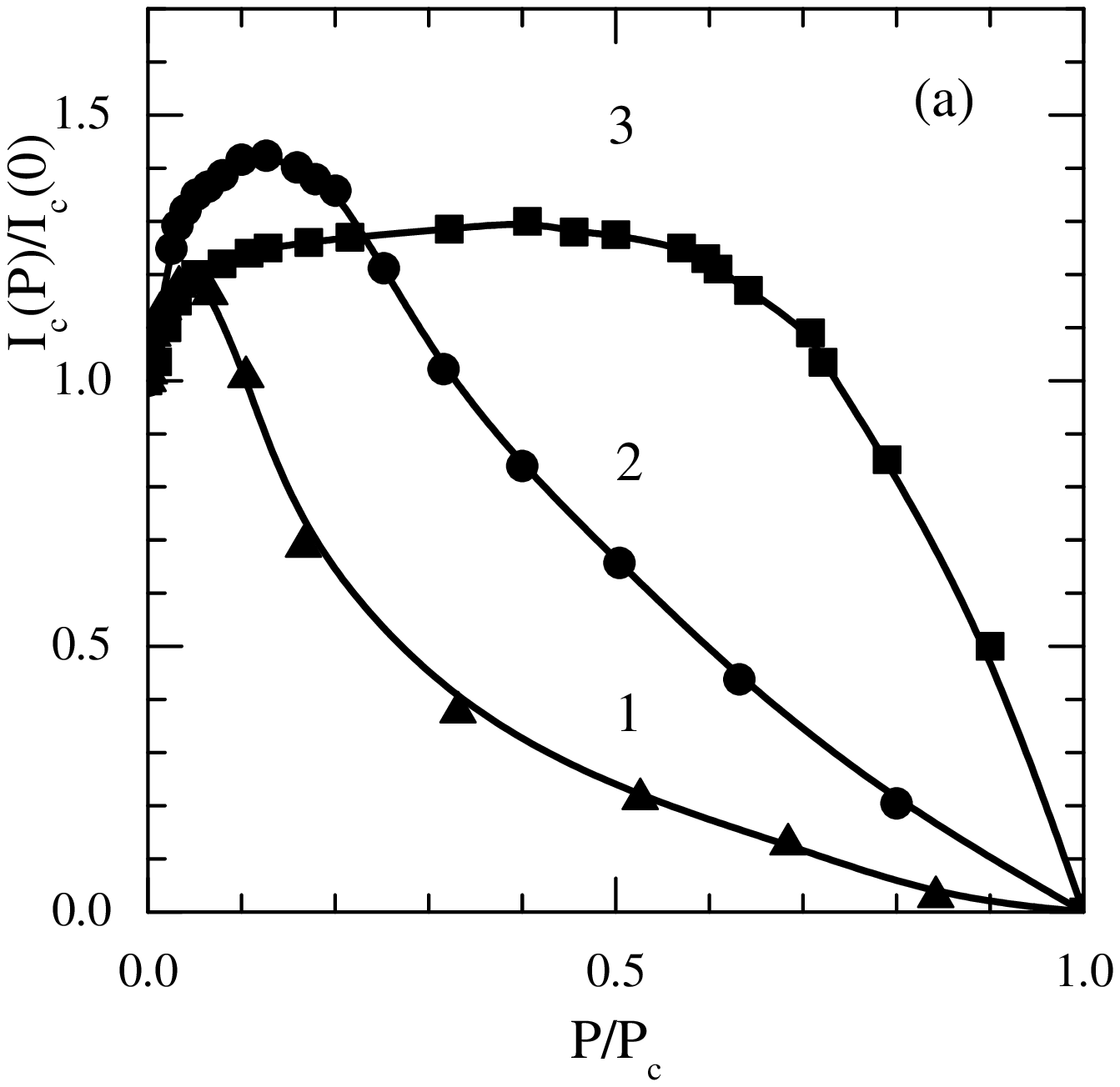}} \vspace{-3mm}
\label{fig3a}\vspace{-3mm}
\end{figure}
\begin{figure}[th]
\centerline{\epsfxsize=4.5cm\epsffile{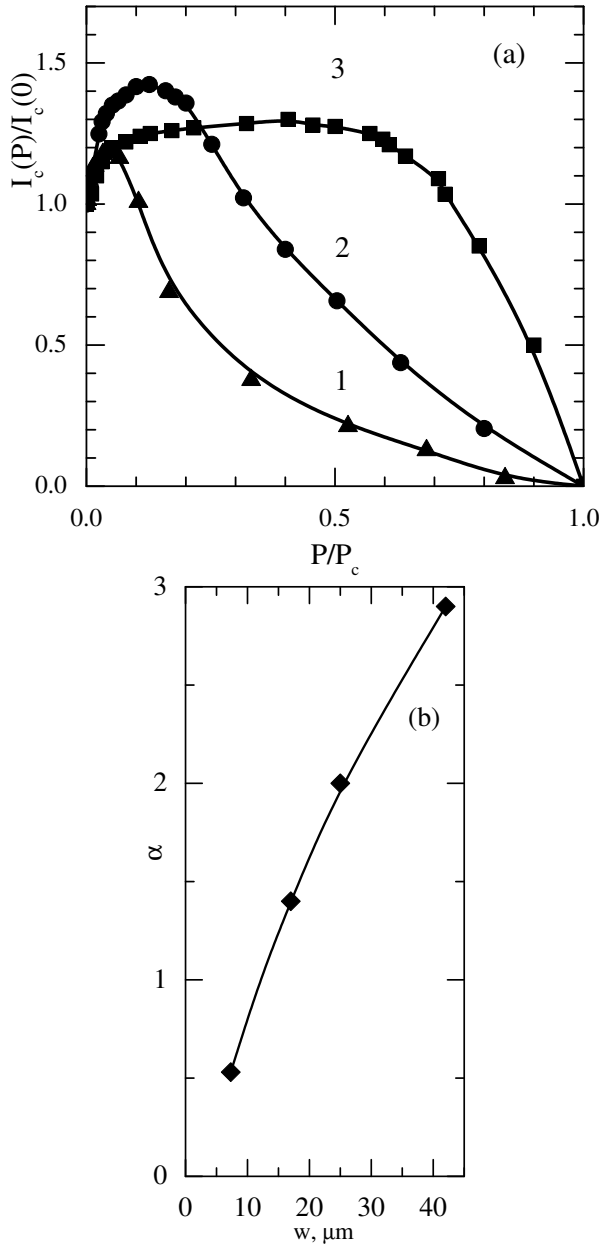}} \vspace{-3mm}
\caption{Relative critical current $I_c(P) / Ic(0)$ versus the reduced
irradiation power $P/P_c$ for different tin films: SnW5 ($\blacktriangle$),
SnW6 ({\Large $\bullet$}), and SnW10 ($\blacksquare$) (a) and the exponent of
its fitting function versus the width $w$ of the sample at 9.2 GHz at
$T/T_c\approx 0.99$ (b).}
\label{fig3b}\vspace{-3mm}
\end{figure}

In summary, the result of the investigations of the critical currents and
different nonequilibrium phenomena in thin films of different width make it
possible to conclude that the widely accepted criterion $w\sim \lam$ for
dividing films into wide and narrow does not work. The transition into the
wide-film regime, described by the existing theory of the vortex state, fully
occurs only for $w/\lam >10-20$. It can be supposed that the simple criterion
$w>\lam$ does not work in practice because of the structural particularities
of a Pearl vortex in a thin film, which is not a truly localized formation
\cite{8}: in contrast to an Abrikosov vortex in a bulk superconductor, its
fields decrease with increasing distance not exponentially but rather in a
power-law fashion. Specifically, Kogan \cite{9} has shown that the magnetic
flux trapped by a single vortex in a thin film, even at a distance $24\lam$
from the vortex core, is 90\% of the flux quantum. The ratio between the film
width and the penetration depth of an electric field, which for $w/\lam \sim
10$ is close to 1, can also play a certain role. The ``anomalous'' vortex
state (transitional region) which we discovered and which arises in the
interval $4<w/\lam <10-20$ for the computed value of the edge current density
$\propto (1-T/T_c)^2$ much smaller than the depairing current density $j_c^
{GL}(T)$ draws attention also. The mechanism of the penetration of vortices
into a film for such small edge currents and the reason why it vanishes for
large values of $w/\lam$ are not described by existing theories and have no
explanation at the present time.


\begin{thebibliography}{99}

\bibitem{1}
I. M. Dmitrenko, Fiz. Nizk. Temp. {\bf 22}, 849 (1996) [Low Temp. Phys. {\bf
22}, 845 (1996)].

\bibitem{2}
L. G. Aslamazov and S. V. Lempitski\v{i}, Zh. Eksp. Teor. Fiz. {\bf 84}, 2216
(1983) [Sov. Phys. JETP {\bf 57}, 1291 (1983)].

\bibitem{3}
A. G. Sivakov, A. P. Zhuravel, O. G. Turutanov, and I. M. Dmitrenko, Czech.
J. Phys. {\bf 46}, 877 (1996).

\bibitem{4}
V. M. Dmitriev and I. V. Zolochevskii, Supercond. Sci. Technol. {\bf 19}, 342
(2006).

\bibitem{5}
G. E. Churilov, V. M. Dmitriev, and A. P. Beskorsy\v{i}, Pis'ma Zh. Eksp.
Teor. Fiz. {\bf 10}, 231 (1969).

\bibitem{6}
G. E. Churidov, V. N. Svetlov, and V. M. Dmitriev, Fiz. Nizk. Temp. {\bf 12},
425 (1986) [Sov. J. Low Temp. Phys. {\bf 12}, 242 (1986)].

\bibitem{7}
V. M. Dmitriev, I. V. Zolochevski\v{i}, T. V. Salenkova, and E. V.
Khristenko, Fiz. Nizk. Temp. {\bf 31}, 1258 (2005) [Low Temp. Phys. {\bf 31},
957 (2005)].

\bibitem{8}
J. Pearl, Appl. Phys. Lett. {\bf 5}, 65 (1964).

\bibitem{9}
V. G. Kogan, Phys. Rev. B {\bf 49}, 15874 (1994)

\end{thebibliography}
\end{document}